\documentclass[aps,prb,reprint,superscriptaddress,amsmath,amssymb,
preprintnumbers,floatfix,frontmatterverbose,nofootinbib,nobibnotes,showkeys]{revtex4-2}
\usepackage{amsmath,amssymb,amsfonts,graphicx,xcolor}
\definecolor{myred}{rgb}{0.843137,0.0980392,0.12549}
\definecolor{myblue}{rgb}{0.121569,0.294118,0.627451}
\definecolor{mygreen}{rgb}{0.0,0.65,0.0}
\definecolor{mygray}{rgb}{0.45,0.45,0.45}
\definecolor{mylightblue}{rgb}{0.187,0.335,0.678}
\definecolor{myorange}{rgb}{0.952,0.423,0.093}
\usepackage[pdfstartview=FitH,colorlinks=true,linkcolor=blue,citecolor=red,urlcolor=mygreen,filecolor=mygray,anchorcolor=myorange,pdfborder={0 0 0}]{hyperref}
\def\EF{E_{\mathrm{F}}}
\def\fbz{$1^{\mathrm{st}}$~BZ}
\begin{document}
\title{Lifshitz-type Reconstruction with Screening-sensitive Fermi-pocket geometry in one-dimensional Metals}
\author{Francesca Mazzei}
\email{anfrace@gmail.com}
\author{Antonello Sindona}
\email{antonello.sindona@fis.unical.it}
\affiliation{\href{https://fisica.unical.it/}{Dipartimento di Fisica}, Universit\`{a} della Calabria, Via P. Bucci, Cubo 30C, I-87036 Rende (CS), Italy\looseness=-1}
\affiliation{\href{https://w3.lnf.infn.it/}{INFN}, Gruppo collegato di Cosenza, Via P. Bucci, Cubo 31C, I-87036 Rende (CS), Italy\looseness=-1}
\begin{abstract}
Electron–electron interactions can alter the Fermi-surface topology of low-dimensional materials beyond conventional bandwidth renormalization.
We predict such a Lifshitz reconstruction in a one-dimensional MoS$_2$ nanoribbon, where $GW$ quasiparticle corrections shift a shallow band minimum below the Fermi level.
The resulting occupied interval splits a previously connected unoccupied region of the lower frontier branch into two symmetry-related hole pockets, adding a pair of Fermi points.
Single-pole and multipole descriptions of dynamical screening agree on this reconstruction while predicting markedly different pocket boundaries and Fermi velocities.
Including electron–hole coupling reveals a screening-dependent redistribution of low-energy spectral weight: the single-pole response is dominated by the metallic continuum near zero energy, whereas the multipole response is dominated by an infrared electron–hole resonance.
\end{abstract}
\maketitle
\section{Introduction}
The low-energy physics of one-dimensional~(1D) conductors is governed by electronic states near a discrete set of Fermi points, whose number and positions constrain the phase space for particle-hole excitations and shape collective charge modes~\cite{haldane1981luttinger, voit1995one}.
When a shallow 1D band extremum lies near the Fermi energy~($\EF$), quasiparticle~(QP) corrections can do more than renormalize bandwidths and group velocities: they can shift the extremum across $\EF$, creating or removing a Fermi-point pair, thereby producing a Lifshitz-type reconstruction of the low-energy electronic structure~\cite{lifshitz1960anomalies}.

Within the $GW$  approximation~\cite{Hedin1965, Godby1989, PhysRevLett.101.226405, Bruneval2014, rohlfing1998electron, Marini2009, Sangalli_2019}, the QP self-energy is built from the dynamically screened Coulomb interaction $W$, whose dependence on the frequency $\omega$ is commonly modeled using single-pole or multipole representations~\cite{larson2013role, soininen2003electron, leon2021frequency, leon2023efficient, guandalini2024efficient}.
Because a Lifshitz reconstruction hinges on QP shifts comparable to the uncorrelated depth of a shallow band extremum, a key question is whether different representations of $W(\omega)$ predict the same change in Fermi-point topology and how the resulting pocket geometry and low-energy response depend on this choice.

Edge-passivated MoS$_2$ nanoribbons provide a natural platform to address this question because chemical edge termination can introduce narrow, edge-derived bands with shallow extrema near $\EF$.
Here, we investigate an N/H-terminated 7-armchair MoS$_2$ nanoribbon~\cite{li2008mos2nanoribbon, ataca2011mos2nanoribbon, pan2012edgedep}, denoted 7AMoS$_2$NR-NH$_2$, using pseudopotential plane-wave density-functional theory~(DFT)~\cite{QE-2009,QE-2017, giannozzi2020quantum, marzari1999thermal, PhysRevB.88.085117, van2018pseudodojo, tantardini2024generating} within the Perdew-Burke-Ernzerhof~(PBE) generalized gradient approximation~(GGA)~\cite{PBE_xc}, supplemented by Grimme's D3 dispersion correction~\cite{grimme2010consistent, grimme2011effect}.
We refine the DFT description of the electronic structure with $GW$ calculations~\cite{Marini2009,Sangalli_2019}, employing the plasmon-pole approximation~(PPA)~\cite{Godby1989, larson2013role} and the multipole approximation~(MPA)~\cite{soininen2003electron, leon2021frequency, leon2023efficient, guandalini2024efficient} as alternative representations of $W(\omega)$.
We show that the GGA electronic structure exhibits two frontier bands, each with one inequivalent Fermi crossing in the irreducible half of the first Brillouin zone~({\fbz}). 

QP corrections shift a shallow minimum of a frontier band below $\EF$, introducing an additional inner crossing near the BZ center~($\Gamma$).
This new crossing and the pre-existing crossing of the other frontier band delimit a hole pocket on the lower frontier branch.
Both screening treatments predict this Lifshitz-type reconstruction but yield markedly different pocket geometries.
This sensitivity to dynamical screening also appears in the low-energy longitudinal response: once electron–hole~(e-h) correlations are included, the PPA spectrum is dominated by the metallic continuum near zero energy, whereas the MPA spectrum is dominated by an infrared e-h resonance.

\section{Fermi-point topology\label{SecCrossing}}
\begin{figure}[h!]
\centering
\includegraphics[width=1.00\linewidth]{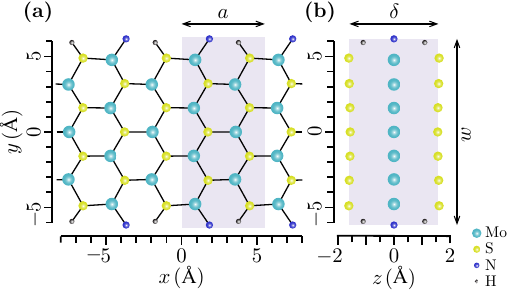}
\caption{\label{geometry}(a) Top and (b) side views of the DFT-optimized geometry of 7AMoS$_2$NR-NH$_2$, with NH$_2$ denoting separate N-Mo and H$_2$-S edge terminations.
The lattice constant is $a=5.515$~{\AA} along the periodic $x$ direction, the width is $w=12.326$~{\AA} along the in-plane $y$ direction, and the out-of-plane span is $\delta=3.214$~{\AA} along the out-of-plane $z$ direction.}
\end{figure}

\begin{figure}[h!]
    \centering
     \includegraphics[width=0.99\linewidth]{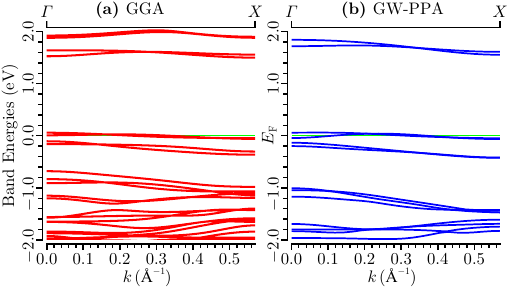}
     \caption{\label{Bands}(a) GGA and (b) GW-PPA band structures of 7AMoS$_2$NR-NH$_2$ along the irreducible $\Gamma X$ segment of the first BZ, with length $\pi/a$~(see Fig.~\ref{geometry}).
     Energies are referenced to the corresponding zero-temperature Fermi level, $\EF=0$.}
\end{figure}

To begin, we consider the relaxed structure of 7AMoS$_2$NR-NH$_2$, periodically repeated along the ribbon axis, as shown in Fig.~\ref{geometry}.
At the armchair edges, nitrogen atoms are bonded to the edge molybdenum atoms, while hydrogen atoms saturate the edge sulfur atoms.
The resulting termination preserves the 1D periodicity of the MoS$_2$ backbone and introduces edge-derived states crossing the Fermi level, yielding a metallic electronic structure.
Further computational details are provided in Appendix~\ref{App:Meths}.
The complete GGA and GW-PPA band structures are shown in Fig.~\ref{Bands}, for wave vectors $0\le k\le 0.5696$~{\AA}$^{-1}$, corresponding to the irreducible $\Gamma X$ segment of the {\fbz}, with edge point $X$. All band dispersions discussed below are presented in this nonnegative $k$-range; their symmetry-related negative-$k$ counterparts are shown in Appendix~\ref{App:GWFullBZ}~(Fig.~\ref{fig:fullBZ}).

QP corrections preserve the metallic character and the overall organization of the electronic manifolds, while generally increasing the separation between occupied and unoccupied states.
A notable feature at both levels of theory is that the bands crossing $\EF$ are separated from the next higher-lying conduction band by at least $1.47$~eV at the GGA level and $1.60$~eV within GW-PPA.
These bands therefore form an energetically isolated frontier manifold that governs the low-energy electronic response.

\begin{figure}[!h]
    \centering
    \includegraphics[width=0.99\linewidth]{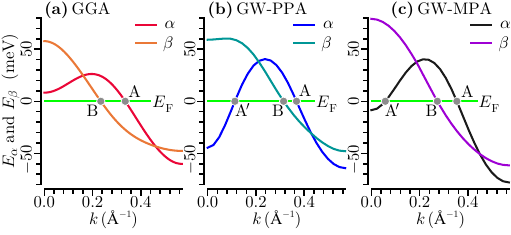}
    \caption{\label{v1BandsZoom}Frontier bands $\alpha$ and $\beta$ of 7AMoS$_2$NR-NH$_2$ calculated within (a)~DFT-GGA, (b)~GW-PPA, and (c)~GW-MPA along ${\Gamma}X$.
    The band energies $E_{\alpha}$ and $E_{\beta}$  follow the continuous dispersions \textit{vs} $k$ through their mutual intersections.
    Circles mark the Fermi crossings, labeled $A$ on $\alpha$ and $B$ on $\beta$ in GGA.
    Both QP calculations introduce an additional crossing $A^{\prime}$ on $\alpha$.
    The positive-$k$ hole pocket is bounded by $A^{\prime}$ and $B$, while $A$ remains the outermost crossing.
    Energies are referenced to the corresponding zero-temperature Fermi level, $\EF=0$, indicated by the horizontal green line.}
\end{figure}

Figure~\ref{v1BandsZoom} compares the frontier-band dispersions $E_\alpha(k)$ and $E_\beta(k)$ obtained within GGA, GW-PPA, and GW-MPA.
The corresponding Fermi wave vectors and absolute band slopes are listed in Table~\ref{tab:fermi}.
The labels $\alpha$ and $\beta$ follow the continuous dispersions through their mutual intersections.
In particular, the lower frontier branch is defined at each $k$ as the lower of $E_\alpha(k)$ and $E_\beta(k)$ and therefore switches between the two labeled dispersions at their intersections.
For each method, the Fermi level is determined by enforcing the same zero-temperature occupation of the interpolated frontier bands, as detailed in Appendix~\ref{App:Meths}.

To compare the two representations of dynamical screening on equal footing, we examine the QP corrections to these two bands, while constructing the screened interaction $W$ from all occupied states and a large number of empty states.
Since $\alpha$ and $\beta$ contain all Fermi-level crossings, their dispersions determine the Fermi-point topology within the present calculation.

Within DFT-GGA~[Fig.~\ref{v1BandsZoom}(a)], $\alpha$ and $\beta$ cross $\EF$ at points $A$ and $B$, respectively, with $k_F^A=0.3355$~\AA$^{-1}$ and $k_F^B=0.2341$~\AA$^{-1}$.
The $\alpha$ band has a shallow local minimum $8.27$~meV above $\EF$ at $\Gamma$, rises to a maximum of $26.2$~meV at $k=0.196$~{\AA}$^{-1}$, and reaches $60.1$~meV below $\EF$ at $X$.
The $\beta$ band decreases from $57.7$~meV above $\EF$ at $\Gamma$ to $47.65$~meV below $\EF$ at $X$.
The corresponding bandwidths are $86.3$ and $105.4$~meV.

Within GW-PPA~[Fig.~\ref{v1BandsZoom}(b)], the $\Gamma$-point minimum of $\alpha$ lies $44.56$~meV below $\EF$, producing an additional crossing point $A^{\prime}$ at $k_F^{A^{\prime}}=0.1127$~\AA$^{-1}$.
The same band reaches a maximum $40.50$~meV above $\EF$ at $k=0.237$~\AA$^{-1}$, crosses $\EF$ again at $A$, with $k_F^A=0.3679$~\AA$^{-1}$, and falls to $64.19$~meV below $\EF$ at $X$.
Its bandwidth therefore increases to $104.7$~meV.
The $\beta$ band lies $59.06$~meV above $\EF$ at $\Gamma$ and has a weak maximum of $60.2$~meV at $k=0.0712$~\AA$^{-1}$ before decreasing through $B$, with  $k_F^B=0.3145$~{\AA}$^{-1}$, to $47.94$~meV below $\EF$ at $X$.
Its bandwidth is $108.2$~meV.

The crossings $A^{\prime}$ on $\alpha$ and $B$ on $\beta$ bound the positive-$k$ hole pocket: the unoccupied interval $k_F^{A^{\prime}}<k<k_F^B$ of the lower frontier branch.
Within this interval, the lower branch follows $\alpha$ up to its intersection with $\beta$ and $\beta$ thereafter.
The crossing $A$ remains the outermost positive-$k$ Fermi point.
In the full BZ, occupation of the interval $-k_F^{A^{\prime}}<k<k_F^{A^{\prime}}$ splits the previously connected unoccupied region of the lower branch into two symmetry-related hole pockets~(see Fig.~\ref{fig:fullBZ}).
The number of inequivalent Fermi points along $\Gamma X$ consequently increases from two in GGA to three in GW-PPA, corresponding to four and six distinct Fermi points in the full BZ.
This change establishes the Lifshitz-type reconstruction.

GW-MPA preserves the three-crossing topology on $\Gamma X$~[Fig.~\ref{v1BandsZoom}(c)] but substantially changes the dispersions.
The $\alpha$ minimum at $\Gamma$ lies $8.53$~meV below $\EF$, and the inner crossing occurs at $k_F^{A^{\prime}}=0.0578$~\AA$^{-1}$.
The band reaches a maximum $40.25$~meV above $\EF$ at $k=0.2136$~\AA$^{-1}$, crosses again at $k_F^A=0.3541$~\AA$^{-1}$, and reaches $77.92$~meV below $\EF$ at $X$.
The $\beta$ band decreases from $79.13$~meV above $\EF$ at $\Gamma$ through $B$, with $k_F^B=0.2734$~\AA$^{-1}$, to $61.5$~meV below $\EF$ at $X$.
The corresponding bandwidths are $118.2$ and $140.6$~meV for $\alpha$ and $\beta$, respectively, approximately $13\%$ and $30\%$ larger than their PPA values.
Both screening models thus place the $\alpha$ minimum below $\EF$, although it is substantially shallower in MPA.

Relative to PPA, MPA moves both boundaries of the $A^{\prime}B$ pocket toward $\Gamma$.
The inner boundary shifts by $0.0549$~\AA$^{-1}$, corresponding to a reduction of approximately $49\%$ in $k_F^{A^{\prime}}$, whereas the outer boundary shifts by $0.0411$~\AA$^{-1}$.
Because the inner boundary moves farther, the pocket width, $\Delta k_{A^{\prime}B}=k_F^B-k_F^{A^{\prime}}$, increases from $0.2017$~\AA$^{-1}$ in PPA to $0.2155$~\AA$^{-1}$ in MPA, a widening of approximately $6.8\%$.
The outermost crossing $A$ shifts toward $\Gamma$ by $0.0138$~\AA$^{-1}$.
\begin{table}[!h]
\centering
\small
\setlength{\tabcolsep}{5pt}
\begin{tabular}{@{}lcccc@{}}
Method & Band & Crossing~($P$) & $k^P_F$ (\AA$^{-1}$) & $\hbar|v^P_F|$ (eV\,\AA) \\
\hline
GGA & $\beta$  & $B$   & 0.234116 & 0.328159 \\
    & $\alpha$ & $A$   & 0.335498 & 0.341069 \\
\hline
GW-PPA & $\alpha$ & $A^{\prime}$ & 0.112715 & 0.569556 \\
       & $\beta$  & $B$          & 0.314452 & 0.339511  \\
       & $\alpha$ & $A$          & 0.367878 & 0.526206 \\
\hline
GW-MPA & $\alpha$ & $A^{\prime}$ & 0.057819 & 0.274306 \\
       & $\beta$  & $B$          & 0.273360 & 0.412011 \\
       & $\alpha$ & $A$          & 0.354073 & 0.549418 \\
\end{tabular}
\caption{\label{tab:fermi}Fermi wave vectors and absolute band slopes of the frontier bands of 7AMoS$_2$NR-NH$_2$ at crossing points $P=A^{\prime},B,A$.}
\end{table}

For the present filling of the two frontier bands with common degeneracy, the zero-temperature occupation constraint requires $S=k_F^A+k_F^B-k_F^{A^{\prime}}$ to equal the length of $\Gamma X$, with $k_F^{A^{\prime}}=0$ in GGA.
The values in Table~\ref{tab:fermi} satisfy this relation within rounding, giving $S\simeq0.5696$~\AA$^{-1}$ for all three methods.
For the QP dispersions, $\Delta k_{A^{\prime}B}=S-k_F^A$, so the pocket widening is matched by the shift of the outermost crossing $A$.

The absolute Fermi slopes, $\hbar|v_F^P|=\left|\partial E/\partial k\right|_{k=k_F^P}$, where $P=A^{\prime},B,A$ labels the crossings, depend nonuniformly on the screening model.
At point $A^{\prime}$, the slope decreases from $0.5696$ to $0.2743$~eV\,{\AA}, a reduction of approximately $52\%$, whereas at $B$ it increases from $0.3395$ to $0.4120$~eV\,{\AA}.
At $A$, the change is smaller, from $0.5262$ to $0.5494$~eV\,{\AA}.
The two screening models therefore agree on the Fermi-point topology but predict different pocket boundary positions and local dispersions, together with a more modest change in pocket width.
These differences modify the phase space and characteristic energy scales of low-energy intraband and interband transitions, providing a microscopic link to the longitudinal response discussed below.

\section{Hole-pocket reconstruction and longitudinal excitations\label{Secabs}}
{\noindent}To examine the consequences of the reconstructed frontier dispersions for longitudinal excitations, we compare the random-phase approximation~(RPA)~\cite{petersilka1996excitation, onida2002electronic, BohmPines1953, HankeSham1975} and Bethe-Salpeter equation~(BSE)~\cite{SalpeterBethe1951, Onida1995} responses within the common transition space spanned by the frontier-band manifold.
Both intraband and interband transitions within $\alpha$ and $\beta$ are included, with occupation factors determined by the corresponding Fermi level at fixed filling.
This comparison examines how explicit e-h correlations modify the response of the frontier manifold.
Transitions involving the remaining bands are examined separately in the $200$-band RPA calculations reported in Appendix~\ref{App:200_TDDFT_RPA}.

Figure~\ref{optspectra} presents the absorptive longitudinal response, $\operatorname{Im}[\epsilon_M(q_{\min},\omega)]$, at the smallest nonzero momentum transfer along the ribbon axis allowed by the adopted $120{\times}1{\times}1$ mesh~\cite{monkhorst1976special}, namely $q_{\min}=9.49\times10^{-3}$~\AA$^{-1}$.
All spectra use a Lorentzian broadening parameter of $0.02$~eV.
The $200$-band RPA calculation retains the strong low-energy QP features and their relative intensity ordering, while modifying their positions and adding finite-energy structure~(Fig.~\ref{appRPAfig}).
It therefore supports the use of the two-band calculation to examine the leading changes in the low-energy RPA response.

In the RPA~[Fig.~\ref{optspectra}(a)], the restricted GGA spectrum is dominated by a feature close to the infrared/terahertz limit~($\omega\to0$), with comparatively weak intensity at higher energies.
The QP dispersions shift the dominant maximum to approximately $0.03$~eV and substantially enhance its peak height.
The PPA and MPA maxima have nearly coincident positions at the adopted broadening, despite their different pocket boundaries and Fermi slopes.
Their weaker structures are more distinct: MPA shows a shoulder in the $0.1$-$0.2$~eV range, whereas PPA produces a different distribution of intensity at higher energies.
These differences reflect the combined changes in transition energies and occupation factors associated with the reconstructed frontier dispersions, including occupation of the interval around $\Gamma$ that separates the two symmetry-related hole pockets.

Including e-h correlations reorganizes the response~[Fig.~\ref{optspectra}(b)].
All three BSE spectra exhibit a finite-energy resonance near $0.12$~eV.
This resonance dominates the GGA and MPA spectra, with similar peak energies and heights.
In PPA, the finite-energy peak is substantially weaker and the largest response occurs close to zero energy.
MPA also retains a low-energy component, but with a markedly smaller peak height than PPA.
The principal distinction between the two QP spectra is therefore the relative prominence of the near-zero-energy structure and the finite-energy resonance.
\begin{figure}[h!]
    \centering
    \includegraphics[width=0.99\linewidth]{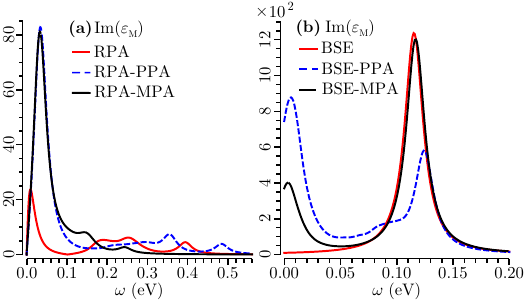}
    \caption{\label{optspectra}Imaginary part of the macroscopic dielectric function, $\operatorname{Im}[\epsilon_M(q_{\min},\omega)]$, calculated within (a)~RPA and (b)~BSE with transitions restricted to the frontier bands $\alpha$ and $\beta$.
    The convention $\hbar=1$ is used, so $\omega$ denotes the excitation energy in eV.
    Both intraband and interband contributions are included.
    Red solid, blue dashed, and black curves denote calculations using GGA, GW-PPA, and GW-MPA energies and occupations, respectively.
    The momentum transfer $q_{\min}=9.49\times10^{-3}$~\AA$^{-1}$ is directed along the ribbon axis.
    All spectra use a Lorentzian broadening parameter of $0.02$~eV.
    The vertical scale in~(b) carries a factor of $10^2$.}
\end{figure}

A finite-energy resonance is thus supported even by the unreconstructed GGA frontier manifold; its appearance alone is not a distinctive signature of the hole-pocket reconstruction.
The QP dispersions instead alter its prominence relative to the low-energy response.
The three calculations use the same transition matrix elements, and the BSE calculations retain the same static e-h interaction matrix elements.
The PPA-MPA differences therefore arise from the QP energies and updated occupation factors, which modify the allowed transitions and their correlated mixing within the same frontier-band manifold.
Within this comparison, the sensitivity to dynamical screening enters the correlated response through the QP input.
We describe the finite-energy structure as a correlation-enhanced e-h resonance.
Agreement on the Fermi-point topology therefore does not imply agreement on the correlated longitudinal response: PPA and MPA predict markedly different relative intensities of the low-energy structure and the infrared e-h resonance.

\section{Summary and conclusion}
{\noindent}We have predicted a QP-driven Lifshitz-type reconstruction of the frontier manifold of 7AMoS$_2$NR-NH$_2$, which converts two positive-$k$ Fermi crossings at the GGA level into three through the formation of two symmetry-related hole pockets in the lower frontier branch.
This topological change is reproduced by both GW-PPA and GW-MPA, whereas the pocket geometry and boundary velocities remain strongly screening dependent.
The resulting differences in carrier phase space persist in the correlated low-energy response, where the two screening treatments produce distinct spectral-weight distributions despite preserving the same Fermi-point topology.
Thus, a robust Fermi-point reconstruction can coexist with substantial screening sensitivity in the pocket geometry, boundary velocities, and correlated low-energy response.

Since these differences occur in the meV-to-sub-eV range, terahertz/infrared spectroscopy could provide experimental access to the predicted screening-sensitive low-energy response, while momentum-resolved photoemission on suitably aligned nanoribbon ensembles could directly probe the reconstructed pocket geometry.
More broadly, our results identify a regime in which shallow bands lying within the energy scale of many-body renormalization can exhibit the same qualitative reconstruction while retaining a pronounced sensitivity of their low-energy properties to dynamical screening.

\section*{Acknowledgments}
\addcontentsline{toc}{section}{Acknowledgments}
The authors acknowledge the use of high-performance computing resources provided by the LEONARDO supercomputing facility, operated by the CINECA consortium (Italy). Computational time was granted under the INF16$\_$npqcd project within the CINECA-INFN agreement and within the INF26\_nemesys project. Additional testing and benchmarking calculations were performed on the Newton high-performance computing cluster at the University of Calabria.

\section*{Funding}
This research was supported by the National Recovery and Resilience Plan (PNRR), funded by the Italian Ministry in Physical, Chemical and Material Sciences and Technologies with Ministerial Decree n$^\textrm{o}$ 351, CUP: H23C22000480007.

\bibliography{MoS2_NH_refs}

\begin{appendix}

\section{Ab~initio Methods\label{App:Meths}}

{\noindent}Equilibrium geometries and Kohn-Sham~(KS) electronic structures were obtained within DFT using the \textsc{Quantum ESPRESSO} suite~\cite{QE-2009,QE-2017,giannozzi2020quantum}.
All GW~\cite{Hedin1965,Godby1989,PhysRevLett.101.226405,Bruneval2014,rohlfing1998electron}, TDDFT-RPA~\cite{petersilka1996excitation,onida2002electronic,BohmPines1953,HankeSham1975}, and BSE~\cite{SalpeterBethe1951,Onida1995} calculations were performed using the \textsc{Yambo} code~\cite{Marini2009,Sangalli_2019}.

The 7AMoS$_2$NR-NH$_2$ system was modeled using a supercell periodic along the ribbon axis, containing seven Mo atoms sandwiched between two S layers of seven atoms each.
The edges were terminated by N atoms bonded to edge Mo atoms and by two H atoms bonded to each terminated edge S atom.
Neighboring periodic images were separated by $20$~{\AA} of vacuum in each transverse direction.
Exchange and correlation were treated within the PBE-GGA~\cite{PBE_xc}, including Grimme's D3 dispersion corrections with zero damping~\cite{grimme2010consistent, grimme2011effect}.
Core-valence interactions were described using nonrelativistic optimized norm-conserving Vanderbilt pseudopotentials~\cite{PhysRevB.88.085117} from the PseudoDojo library~\cite{van2018pseudodojo,tantardini2024generating}.
The KS orbitals and charge density were expanded in plane waves with kinetic-energy cutoffs of $92$ and $400$~Ry, respectively.
Atomic coordinates and the lattice parameter along the periodic direction were optimized using the Broyden--Fletcher--Goldfarb--Shanno algorithm until all Cartesian force components were below $10^{-5}$~Ry/Bohr and the total-energy change between consecutive ionic steps was smaller than $10^{-12}$~Ry.
The optimized structure is shown in Fig.~\ref{geometry}.
The ground-state charge density was calculated using a $120{\times}1{\times}1$ Monkhorst--Pack grid~\cite{monkhorst1976special}, Marzari--Vanderbilt cold smearing~\cite{marzari1999thermal} with a width of $0.01$~Ry, and an electronic self-consistency threshold of $10^{-12}$~Ry.

The $G_0W_0$ calculations employed $600$ KS bands and a $48{\times}1{\times}1$ BZ grid.
The exchange contribution to the self-energy was evaluated with a reciprocal-space cutoff of $40$~Ry.
The correlation contribution used all $600$ bands and a response-block cutoff of $12$~Ry.
Matrix elements of the exchange-correlation potential entering the QP corrections were evaluated with a cutoff of $160$~Ry.
The frequency dependence of the screened interaction was treated either within PPA~\cite{Godby1989}, using an imaginary reference energy of $1$~Ha, or within MPA~\cite{soininen2003electron,leon2021frequency,leon2023efficient}, using $16$ poles.
QP corrections to the frontier bands $\alpha$ and $\beta$ were interpolated within \textsc{Yambo} from the $48{\times}1{\times}1$ grid onto the $120{\times}1{\times}1$ grid used for the response calculations.

For the band analysis and dielectric response settings, $\EF$ was calibrated independently at zero temperature within either GGA, GW-PPA, or GW-MPA, keeping the electron number fixed at $196$ per unit cell.
Since the remaining bands are fully occupied or empty, this condition is equivalent to retaining two electrons in the spin-degenerate frontier manifold.
The Fermi level was obtained from the continuous interpolated dispersions by solving
\begin{equation}
\sum_{\nu=\alpha,\beta}\int_0^{\pi/a}
\Theta\!\left[\EF-E_\nu(k)\right]\,dk=\frac{\pi}{a},
\label{eq:SI_fermi_calibration}
\end{equation}
where $\Theta$ is the Heaviside function and $\pi/a$ is the length
of the irreducible $\Gamma X$ segment of the {\fbz}~[Figs.~\ref{Bands} and~\ref{v1BandsZoom}].
This fixes the total occupied one-dimensional $k$-space measure and implies
$k_F^A+k_F^B-k_F^{A^{\prime}}=\frac{\pi}{a}$, with $k_F^{A^{\prime}}=0$ in GGA.
The calibrated Fermi levels define the energy references in the band plots and the occupation factors entering the response calculations.
This zero-temperature calibration is separate from the cold smearing used to converge the ground-state density.

The longitudinal TDDFT-RPA and BSE responses were evaluated on a $120{\times}1{\times}1$ grid at the smallest nonzero momentum transfer,
$q_{\min}=2\pi/(120a)=9.49\times10^{-3}$~\AA$^{-1}$, directed along the ribbon axis.
The $200$-band RPA calculations included more than $100$ unoccupied bands and states extending approximately $10$~eV above $\EF$.
The statically screened interaction entering the BSE kernel was constructed using the same $200$-band manifold.
The bare-exchange and screened-direct components of the BSE kernel were evaluated with reciprocal-space cutoffs of $45$ and $4$~Ry, respectively.
To isolate the response associated with the reconstructed frontier
manifold and enable a direct RPA-BSE comparison, the main-text
calculations were restricted to intraband and interband transitions
within the $\alpha$ and $\beta$ bands.
The same KS transition matrix elements were used in GGA, GW-PPA, and GW-MPA, and the BSE calculations retained the same static e-h interaction matrix elements.
Only the frontier-band energies and their corresponding occupation factors were varied.
Thus, the screening-model dependence of this comparison enters through the QP input, while the static e-h interaction matrix elements remain fixed.
All spectra were evaluated using a Lorentzian broadening parameter of $0.02$~eV.

\section{GW-PPA and GW-MPA reconstruction on the full BZ\label{App:GWFullBZ}}
\begin{figure}[h!]
\centering
\includegraphics[width=0.9\linewidth]{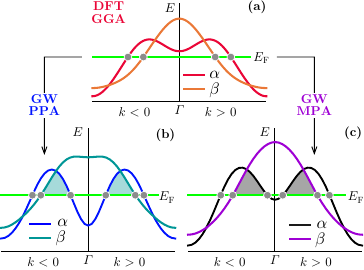}
\caption{\label{fig:fullBZ}
Schematic reconstruction of the frontier $\alpha$ and $\beta$  bands of 7AMoS$_2$NR-NH$_2$ across the full BZ: (a) DFT-GGA, (b) $GW$-PPA, and (c) $GW$-MPA.
Both $GW$ treatments place the $\Gamma$-centered minimum of $\alpha$ below $\EF$, splitting its connected unoccupied region into two symmetry-related hole pockets and increasing the total number of Fermi points from four to six.
Green lines indicate the corresponding Fermi levels, gray circles mark Fermi points, and shading highlights the hole pockets.
}
\end{figure}

Figure~\ref{fig:fullBZ} illustrates schematically how the two frontier branches of 7AMoS$_2$NR-NH$_2$ are reconstructed by $GW$ quasiparticle corrections across the full $-\pi/a \leq k \leq \pi/a$ range, which delimits the {\fbz}.
The dispersions are symmetric under $k\rightarrow -k$, with $\Gamma$ at $k=0$, so Fermi points occur in symmetry-related pairs.
For each electronic structure, occupations are defined relative to the corresponding Fermi level, determined at the same total filling.

Within DFT-GGA, the shallow local minimum of $\alpha$ at $\Gamma$ lies above $\EF$.
The unoccupied region of the lower frontier branch therefore extends continuously across $\Gamma$ and is bounded by the two symmetry-related $B$ crossings.
Together with the two $A$ crossings on $\alpha$, this gives four Fermi points across the full BZ.

Both the GW-PPA and the GW-MPA approaches to dynamical screening place the $\Gamma$-centered minimum of $\alpha$ below $\EF$.
An occupied interval consequently appears around $\Gamma$, splitting the previously connected unoccupied region of the lower frontier branch into two symmetry-related hole pockets in the intervals
$k_F^{A^{\prime}}<|k|<k_F^{B}$.
The $\alpha$ band thus contributes four Fermi points, while $\beta$ retains two.
The total number increases from four in GGA to six in both $GW$ descriptions, establishing the Lifshitz character of the reconstruction.

The two screening approximations nevertheless yield different reconstructed dispersions.
The PPA minimum lies farther below $E_{\mathrm F}$, whereas the MPA minimum remains closer to it.
The resulting differences in the inner and outer Fermi wave vectors determine distinct hole-pocket widths, $\Delta k_{A^{\prime}B}$.
Differences in the band slopes at the crossings likewise produce different Fermi velocities.
The full-BZ representation therefore highlights a common change in Fermi-point topology together with a screening-dependent geometry of the reconstructed pockets.

\section{$200$-band TDDFT-RPA response\label{App:200_TDDFT_RPA}}
\begin{figure}[h!]
\includegraphics[width=0.99\linewidth]{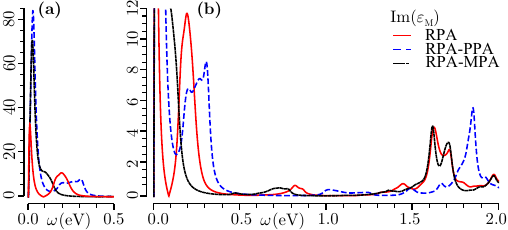}
\caption{\label{appRPAfig}Imaginary part of the macroscopic dielectric function,
$\operatorname{Im}[\epsilon_M(q_{\min},\omega)]$, calculated within RPA using $200$ bands:
(a) low-energy range up to $0.5$~eV on the full intensity scale;
(b) extended range up to $2.0$~eV with a reduced vertical range
that reveals weaker structures and clips the dominant low-energy peaks.
As in Fig.~\ref{optspectra}, $\omega$ denotes the excitation energy in eV.
Red solid, blue dashed, and black short-dashed curves denote GGA, GW-PPA, and GW-MPA results, respectively.
Only the frontier bands $\alpha$ and $\beta$ have QP-corrected energies and updated occupations in the latter two calculations.
The momentum transfer $q_{\min}=9.49\times10^{-3}$~\AA$^{-1}$ is along the ribbon axis, and the Lorentzian broadening parameter is $0.02$~eV.}
\end{figure}

The main-text RPA-BSE comparison uses transitions within the same frontier-band manifold, $\alpha$ and $\beta$.
To assess the contributions of the remaining bands at the RPA level, Fig.~\ref{appRPAfig} presents spectra obtained using $200$ bands.
The momentum transfer, BZ sampling, KS transition matrix elements, and spectral broadening are unchanged.
In the RPA-PPA and RPA-MPA calculations, only the energies and occupations of $\alpha$ and $\beta$ were updated; all other bands retained their GGA energies.

The dominant low-energy response is retained upon inclusion of the  $200$-band manifold~[Fig.~\ref{appRPAfig}(a)].
The PPA maximum remains higher than the MPA maximum, and both
exceed the corresponding GGA maximum.
The PPA and MPA peak energies lie in the $0.02$-$0.03$~eV range,
compared with approximately $0.03$~eV in the two-band calculation.
Additional transitions modify the finite-energy structure, including a pronounced GGA feature near $0.19$~eV and PPA structures in the $0.2$-$0.4$~eV range.
The two-band calculation in Fig.~\ref{optspectra}(a) therefore captures the leading enhancement of the low-energy RPA peak heights and their relative ordering, while the additional bands affect the detailed spectral structure.

The extended energy range~[Fig.~\ref{appRPAfig}(b)] reveals weaker structures associated with the enlarged transition manifold.
The most prominent higher-energy features occur in the
$1.6$-$1.9$~eV range.
GGA and MPA exhibit similar structures between $1.6$ and
$1.75$~eV, whereas the dominant PPA feature lies near $1.85$~eV.
Because all other band energies are unchanged, these differences
arise from the modified frontier-band energies and occupation
factors and their effect on the RPA screening.
This comparison supports the two-band description of the leading low-energy RPA features.
\end{appendix}

\end{document}